\documentclass[12pt]{article}
\usepackage{xcolor}    
\usepackage{amssymb}   
\usepackage{listings}
\usepackage{graphicx}
\usepackage{amsmath, amssymb}
\usepackage{authblk}
\usepackage{geometry}
\usepackage[colorlinks=true, linkcolor=blue, citecolor=blue, urlcolor=blue]{hyperref}

\title{Preliminary Design and Performance Simulation of a Thermal Neutron Diffractometer Using McStas}
\author[1, 2]{Li-Fang Chen\thanks{Corresponding author: lifangchen0507@gmail.com}}

\affil[1]{Department of Physics, National Taiwan Normal University, Taipei, Taiwan}
\affil[2]{National Atomic Research Institute (NARI), Taoyuan, Taiwan}

\date{}
\usepackage{multirow}
\usepackage{adjustbox} 

\begin{document}
\maketitle

\begin{abstract}
This study presents a preliminary design and simulation of a thermal neutron diffractometer using the McStas Monte Carlo ray-tracing package. The simulated system is based on generalized instrument parameters, including typical collimator divergences, monochromator mosaic spreads, and detector positioning, rather than on any specific engineering blueprint. The simulation evaluates key performance metrics such as neutron flux distribution, beam divergence, and wavelength resolution along the beam path.

Results demonstrate that, even under simplified geometric assumptions, the model can provide insight into component alignment, resolution limitations, and flux attenuation behavior. The modular structure of the simulation allows easy parameter adjustments for different instrument configurations and serves as a foundation for more advanced system-specific optimization studies.

Additionally, the provided McStas input file serves as a reusable and customizable simulation template that can significantly reduce modeling time for engineering personnel during the early-stage design and layout phase. This tool also facilitates technical training and comparative analysis without relying on finalized engineering drawings.


\end{abstract}

\maketitle

\section{Introduction to Neutron Diffractometers and McStas Software}
The fundamental principle of a neutron diffractometer is based on the wave properties of neutrons and their interaction with the periodic arrangement of atoms in a crystal. When incident neutrons interact with a crystal, diffraction occurs, altering the propagation direction of the neutrons within the crystal and producing a diffraction pattern. According to Bragg's law, there is a specific relationship between the diffraction angle and the lattice constant, which allows for the determination of the crystal structure of a material by analyzing the diffraction pattern.
McStas is a software package designed for performing Monte Carlo simulations of complex neutron scattering instruments\cite{lefmann1999, willendrup2022}. It is available for use on Windows, Macintosh, and UNIX/Linux platforms. McStas integrates with MATLAB for 3D visualization of instrument geometries. As a versatile neutron ray-tracing simulation tool, McStas provides accurate estimates of flux, resolution, parameter optimization, and design, especially in cases where analytical calculations are not feasible.
Initially developed in 1997 at Risø National Laboratory in Denmark, McStas has since been adopted by institutions such as the Institut Laue-Langevin (ILL) in France and ISIS Neutron and Muon Source in the UK. It is based on a domain-specific language designed for neutron scattering, which is efficiently translated into ANSI-C and then into executable files for simulations. The domain-specific language allows users to construct instruments from individual components, drawing from a standard component library maintained by the user community. Each component is programmed in C++ to simulate the corresponding physical part, collectively forming a neutron scattering instrument. Notable components include moderators, guides, samples, analyzer crystals, and detectors. Users can also design custom components for use within their instruments.
One key advantage of neutron ray-tracing simulations is the ability to place monitors at any desired location without disturbing the beam. Recent versions of McStas also support Python scripting through the mcstas-script interface, enabling automated simulation workflows and facilitating post-processing and data analysis. The version of McStas used in this work is 3.2.

\section{Layout of Thermal Neutron Diffractometer Components and Input Interface Design in McStas}
\begin{figure}[htbp]
  \centering
  \includegraphics[width=0.85\textwidth]{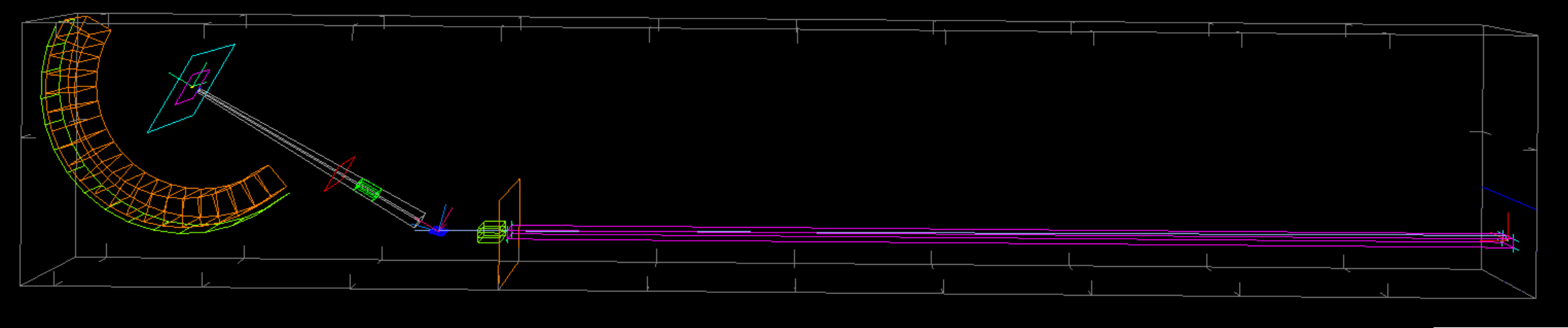}
  \caption{Component layout of the thermal neutron diffractometer simulated using McStas. The simulation includes major optical components such as the neutron guide, collimators, monochromator, sample stage, and detector.}
  \label{fig:layout}
\end{figure}

The component layout of the thermal neutron diffractometer is illustrated in Figure~\ref{fig:layout}. A neutron guide is positioned 2~cm downstream from the neutron source, with a default entrance cross-section of 6~cm~$\times$~12~cm and a total length of 7~meters. All related parameters can be modified by users through the provided input file. Adjustable slits are placed at both the entrance and exit of the guide to control the beam cross-sectional profile. The \texttt{McStas} simulation begins at the entrance of the neutron guide and includes all major optical components up to the detector. The geometric layout of each component is shown in its corresponding position in Figure~\ref{fig:layout}.

\begin{figure}[htbp]
  \centering
  \includegraphics[width=0.85\textwidth]{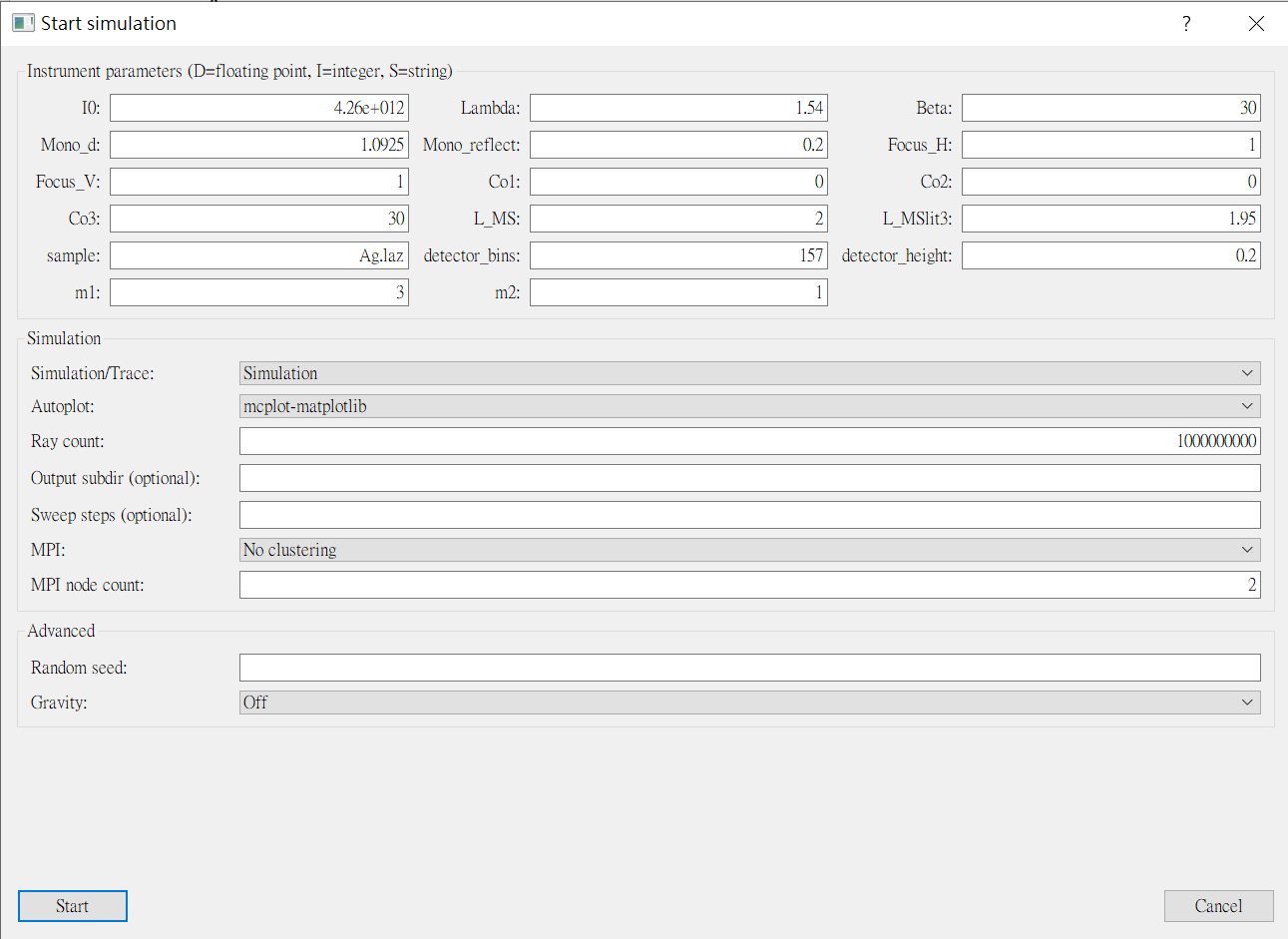}
  \caption{Graphical interface for McStas input parameters.
The interface includes 17 adjustable fields, covering neutron wavelength, monochromator settings (mosaic spread, d-spacing, reflectivity, focusing options), collimator angles, sample type, detector size, and neutron guide m-value. Default values are preloaded and can be modified as needed. See Appendix A for details.
}
  \label{fig:interface}
\end{figure}

Figure~\ref{fig:interface} presents the parameterized input interface developed in this study using \texttt{McStas}. The interface includes 17 adjustable fields that cover the neutron wavelength at the sample position, as well as various settings related to the single-crystal monochromator—such as mosaic spread, interplanar spacing, reflectivity, and switches for vertical and horizontal focusing. Additional adjustable parameters include the geometry of the three collimators, distances between key components (e.g., from the monochromator to the sample and from the monochromator to the slit), sample type, detector pixel size and height, and the \textit{m}-value of the neutron guide. When the input file is loaded, the interface automatically populates the default values. A detailed explanation of all adjustable fields is provided in \hyperref[appendix:A]{\textbf{Appendix A}}.

\section{Overview of Key Components in the Neutron Diffractometer and Their Corresponding Modules in McStas}

\subsection{Neutron Source}

McStas provides a variety of neutron source modules that can be tailored to specific simulation needs. In this study, a representative thermal neutron source is adopted using the \texttt{Source\_Maxwell\_3} module, which simulates a Maxwell-Boltzmann distribution of neutron energies suitable for thermal neutron instruments.

This module allows users to define the characteristic temperature \texttt{TTT} (in Kelvin), which governs the energy distribution of emitted neutrons. Additional configurable parameters include the source position, emission angles, and directionality. All of these can be adjusted within the \texttt{Source\_Maxwell\_3} section of the McStas input file.

By editing the parameters associated with this component, users can customize the simulation to reflect various neutron source characteristics and evaluate the corresponding effects on instrument performance.

\subsection{Neutron Guide}

In this simulation, the neutron guide is implemented using the \texttt{Guide} component provided in the McStas library. This module models a rectangular guide aligned along the Z-axis (the beam direction), with its entrance face lying in the X-Y plane.

The neutron guide configuration consists of two primary aspects: geometric dimensions and internal coating properties.

In the current simulation setup, the guide is 700~cm in length, with both entrance and exit cross-sectional dimensions set to 6~cm (width) $\times$ 12~cm (height). The interior of the guide is assumed to be under vacuum. These geometric parameters can be modified in the \texttt{Guide} module to simulate different beam transport conditions and guide shapes.

The reflective properties of the guide are defined by the \texttt{m}-value, which characterizes the surface coating’s critical angle for total reflection. An \texttt{m = 0} corresponds to a perfectly absorbing wall, \texttt{m = 1} simulates standard nickel-coated surfaces, and \texttt{m > 1} indicates the use of supermirror coatings that allow reflection of neutrons with higher incident angles. In practical terms, an \texttt{m = 3} supermirror coating has a critical reflection angle of approximately 0.49$^\circ$ for neutrons with a wavelength of 1.65~\AA---closely matching the horizontal divergence determined by the guide’s aspect ratio (700~cm in length and 6~cm in width).

Users may adjust the \texttt{Guide} component parameters---including length, cross-sectional area, and \texttt{m}-value---to study how different guide designs affect neutron transport and beam divergence.

\subsection{Monochromator}

In McStas simulations, the function of a monochromator is implemented using the \texttt{Monochromator\_curved} component. This module models a curved, mosaic single crystal that is widely used for extracting monochromatic neutron beams from a polychromatic source.

The working principle is based on Bragg's law: when neutrons strike a crystal at an angle $\theta$, only wavelengths satisfying the condition $\lambda = 2d \sin \theta$ will be constructively reflected. By selecting appropriate interplanar spacing $d$ and incident angle $\theta$, one can effectively isolate neutrons of a specific wavelength. This principle is extensively applied in the design of neutron diffraction and scattering instruments.

In this simulation, the monochromator is constructed from a $5 \times 5$ array of Highly Oriented Pyrolytic Graphite (HOPG) crystals, with each crystal plate measuring 2~cm $\times$ 3~cm and spaced 0.05~mm apart. The component supports both vertical and horizontal focusing, which can be toggled through input parameters. Users may also define the curvature radius and mosaic spread of the crystal assembly.

Regarding reflectivity settings, the HOPG(002) reflectivity data can be directly loaded from McStas’s internal database. If using a different crystal such as Ge(115), the reflectivity and transmission profiles must be specified manually.

Adjustable parameters in the \texttt{Monochromator\_curved} module include:
\begin{itemize}
    \item Crystal material and reflectivity data file
    \item Crystal dimensions, quantity, and spacing
    \item Mosaic spread
    \item Vertical and horizontal focusing options
    \item Interplanar spacing $d$
\end{itemize}

These options provide users with flexible modeling capabilities to explore how various monochromator designs affect neutron wavelength selection and beam focusing characteristics.

\subsection{Collimator}

A collimator is an optical component designed to control the divergence of a neutron beam. Its primary function is to confine a divergent neutron flux within a defined angular range, thus producing a quasi-parallel beam. This is essential for improving the resolution and signal-to-noise ratio in diffraction experiments by minimizing background and stray neutrons.

In this study, two types of McStas collimator components are used:
\begin{itemize}
    \item \texttt{Collimator\_linear}: This component simulates a linear collimator and is used at two positions—between the neutron guide and the monochromator (Collimator 1), and between the monochromator and the sample (Collimator 2). Users can specify opening dimensions, length, and angular range to restrict neutron passage.
    \item \texttt{Collimator\_radial}: This component simulates a radial collimator commonly used around the sample position in powder diffractometers. It is placed between the sample and the detector. Adjustable parameters include the number of channels, angular aperture, length, and shielding material.
\end{itemize}

These collimator components allow users to fine-tune angular constraints and spatial definitions of each neutron beam section via the input file, offering high flexibility in simulating various diffraction geometries.

Key adjustable parameters include:
\begin{itemize}
    \item Aperture width, height, and length
    \item Maximum allowed divergence angle (angular acceptance)
    \item Shielding material and absorption probability (optional)
\end{itemize}

With these settings, users can better capture the influence of collimation on flux and resolution, supporting instrument optimization and experimental planning.

\subsection{Sample Stage}

The sample stage is a critical structural component of a neutron diffractometer. It supports and positions the experimental sample while maintaining the required geometry to align with the neutron beam emerging from the monochromator. In this design, the sample stage is fabricated from 6061 aluminum alloy and includes both vertical adjustment and rotational functionality. The rotation radius is approximately 2 meters, centered on the monochromator, allowing for flexibility in accommodating different sample types and take-off angles.

A neutron guide connects the monochromator to the sample stage, with a slit positioned at the guide’s outlet near the sample to restrict the beam cross-section. This slit, with a width of 2~cm and a height of 3~cm, enhances beam collimation and suppresses stray neutrons, improving the incident beam quality.

To simplify the simulation process and focus on diffraction phenomena, structural details of the aluminum sample stage and sample holder are omitted in the McStas model. Instead, the \texttt{PowderN} component is used to simulate the elastic scattering behavior of polycrystalline samples, suitable for powder diffraction studies. The sample material can be adjusted via input parameters, with the default set to silver powder.

This modeling simplification enables efficient simulation of diffraction intensity distributions and resolution effects, minimizing interference from structural supports or sample containers. It also provides flexibility for varying sample types and geometries, supporting iterative instrument design and comparative analysis.

\subsection{Detector}

In this simulation, the detector is implemented using the \texttt{Monitor\_nD} component from the McStas library, positioned downstream of the sample stage. Its role is to record the spatial distribution of neutrons scattered by the sample. To evaluate the angular resolution and spatial performance of the thermal neutron diffractometer, the \texttt{Monitor\_nD} is configured to capture neutron intensity over a scattering angle range of 3$^\circ$ to 160$^\circ$. The output includes both one-dimensional and two-dimensional data formats.

The effective detector height is set to 20~cm, covering the main angular region behind the sample stage. As real neutron detectors exhibit efficiency variations based on operating voltage and readout electronics, this simulation assumes an ideal detector with 100\% efficiency—neglecting neutron losses and detector response imperfections. This idealization allows the study to focus on flux behavior and geometric effects on diffraction patterns, facilitating clean performance comparisons.

\section{Use Case Example}

To demonstrate the practicality of the developed McStas input framework, this section presents a representative simulation case and corresponding analysis. The example simulates the diffraction pattern of a thermal neutron powder diffractometer using the customized input file developed in this study.

Each diffraction peak in the simulated spectrum is fitted using a Gaussian function to extract its full width at half maximum (FWHM). These FWHM values are then converted into lattice resolutions based on the formulation proposed by A.~W.~Hewat in 1975~\cite{hewat1975}. 

The resulting lattice resolution values are further compared with the theoretical predictions derived from the analytical formula established by G.~Caglioti et al. in 1958~\cite{caglioti1958}. This comparative analysis serves as a verification of the simulation tool's accuracy and demonstrates its applicability in early-stage neutron instrument design and resolution assessment.

\subsection*{Example: Neutron Flux and Lattice Resolution Using a Ge(115) Monochromator at a Take-Off Angle of 89.63$^\circ$}

The lattice spacing of the Ge(115) crystal plane is 1.0925~\AA, making it a suitable choice for monochromator applications under high take-off angle configurations while still covering the target neutron wavelength range in this study. A mosaic spread of 0.5$^\circ$ was selected to achieve a balance between resolution and intensity. According to the work by T.~Kittelmann \textit{et al.}\cite{kittelmann2021}, under the conditions of a 1~cm thick crystal, 0.5$^\circ$ mosaic spread, and a 90$^\circ$ take-off angle, the reflectivity of Ge(115) is approximately 0.2. Based on these parameters, a neutron wavelength of 1.54~\AA{} was selected for sample illumination.

Figure 3 presents the simulated powder diffraction pattern for a silver sample under the specified configuration. Table 1 summarizes the diffraction peak data obtained from McStas simulations under three monochromator focusing conditions: no focusing, vertical focusing, and double-axis focusing. Each diffraction peak was fitted using a Gaussian function, and the corresponding lattice resolutions were calculated. A comparison between the simulation results and the analytical resolution predictions is provided in Figure 4, demonstrating good agreement under the no-focusing condition.

It should be noted that the horizontal axis in Figure 3 displays diffraction angles ranging from \(-160^\circ\) to \(-5^\circ\). This setting arises from the "parallel" detector configuration commonly used in diffractometers to achieve higher resolution performance\cite{caglioti1958}. Users are encouraged to modify the "Monitor\_nD()" component settings in the input file—specifically, by adjusting the \texttt{theta\_limits} parameter to \([5, 160]\)—in order to compare results under different angular detection ranges.

\begin{figure}[htbp]
\centering
\includegraphics[width=0.8\textwidth]{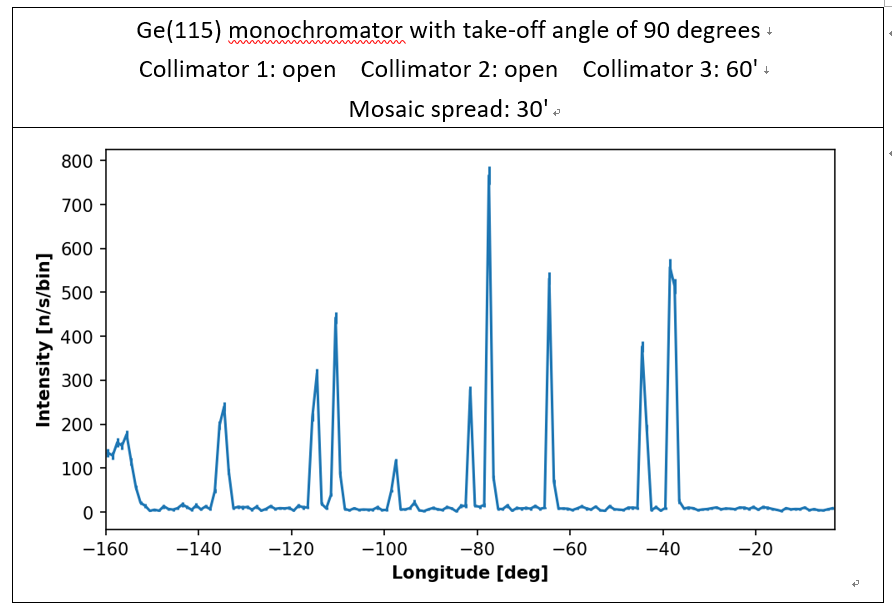} 
\caption{Simulated neutron powder diffraction pattern for a silver sample using Ge(115) as the monochromator crystal. The configuration includes a take-off angle of 90$^\circ$, a mosaic spread of 30$'$, Collimator 1 and 2 in open state, and Collimator 3 set to 60$'$. The figure shows the angular distribution and relative intensity of scattered neutrons within the 3$^\circ$ to 160$^\circ$ detection range. Multiple well-resolved diffraction peaks are observed, providing a solid basis for subsequent resolution analysis and theoretical comparison.}
\label{fig:diffraction_pattern}
\end{figure}

\begin{figure}[htbp]
    \centering
    \includegraphics[width=0.8\textwidth]{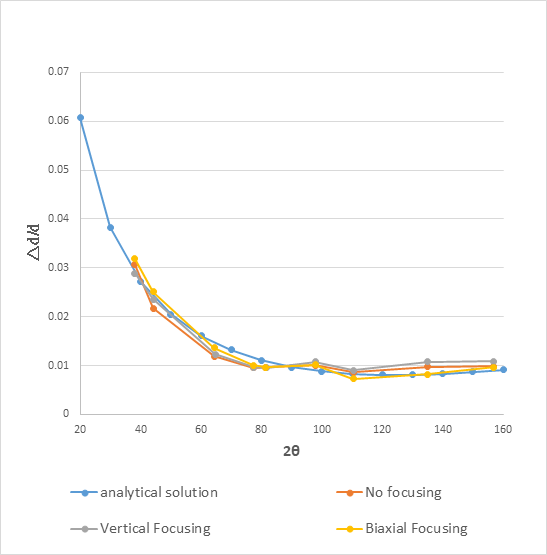} 
    \caption{Comparison between McStas-simulated and analytical diffraction resolutions. Simulated lattice resolutions under three monochromator focusing modes---no focusing, vertical, and biaxial---are compared with the analytical formula by Caglioti et al.\ (1958). The no focusing case shows excellent agreement with theory.}
    \label{fig:resolution_comparison}
\end{figure}

\begin{table}[htbp]
\centering
\caption{Simulated neutron flux and diffraction resolution at the sample position using a Ge(115) monochromator with a 90° take-off angle, based on McStas simulations.}
\label{tab:flux_resolution}
\resizebox{\textwidth}{!}{ 
\begin{tabular}{|c|cc|cc|cc|}
\hline
\multicolumn{7}{|c|}{Ge(115) monochromator with take-off angle of 90 degrees} \\ 
\multicolumn{7}{|c|}{Collimator 1: open \quad Collimator 2: open \quad Collimator 3: 60'} \\
\multicolumn{7}{|c|}{Mosaic spread: 30'} \\ \hline
\multirow{10}{*}{Lattice Resolution} 
& \multicolumn{2}{c|}{No Focusing} 
& \multicolumn{2}{c|}{Vertical Focusing} 
& \multicolumn{2}{c|}{Biaxial Focusing} \\ \cline{2-7}
& Diffraction Angle (°) & $\Delta d/d$ 
& Diffraction Angle (°) & $\Delta d/d$ 
& Diffraction Angle (°) & $\Delta d/d$ \\ \cline{2-7}
& 38.01 & 0.0306 & 37.99 & 0.0289 & 38.01 & 0.0319 \\ \cline{2-7}
& 44.21 & 0.0217 & 44.25 & 0.0236 & 44.20 & 0.0251 \\ \cline{2-7}
& 64.39 & 0.0119 & 64.66 & 0.0123 & 64.35 & 0.0136 \\ \cline{2-7}
& 77.38 & 0.0095 & 77.38 & 0.0096 & 77.30 & 0.0101 \\ \cline{2-7}
& 81.40 & 0.0096 & 81.52 & 0.0095 & 81.44 & 0.0096 \\ \cline{2-7}
& 97.86 & 0.0101 & 97.75 & 0.0108 & 97.80 & 0.0102 \\ \cline{2-7}
& 110.39 & 0.0086 & 110.41 & 0.0090 & 110.44 & 0.0073 \\ \cline{2-7}
& 134.82 & 0.0098 & 134.86 & 0.0108 & 134.80 & 0.0082 \\ \cline{2-7}
& 156.80 & 0.0099 & 156.74 & 0.0109 & 156.83 & 0.0097 \\ \hline
\end{tabular}
} 
\end{table}

It should be noted that, as shown in Table~\ref{tab:flux_resolution} and Figure~\ref{fig:resolution_comparison}, the effects of focusing conditions on the lattice resolution appear rather limited, which is clearly counterintuitive from a physical standpoint. In reality, both vertical and biaxial focusing have a significant impact on resolution, with biaxial focusing exerting a particularly pronounced effect.

The reason why this degradation is not captured through the Gaussian fitting of simulated diffraction peaks is that biaxial focusing substantially alters the wavelength distribution of neutrons reaching the sample stage, leading to a shift in the peak position of the wavelength spectrum. This phenomenon can be verified by examining the output file \texttt{E\_Monitor\_sample\_3.txt}. As a result of the modified wavelength distribution, the positions of the diffraction peaks are also displaced.

Although this deterioration in lattice resolution cannot be directly observed through conventional Gaussian fitting analysis, it will in practice impair the accuracy of subsequent crystal structure refinements based on the diffraction data.

\section{Discussion and Customization Potential}

This section discusses the flexibility and extensibility of the McStas input architecture developed in this study. The modular simulation structure is designed to support future use cases in instrument design and parameter space exploration. Key advantages include:

\begin{enumerate}
    \item \textbf{Modular Parameter Design} \\
    The input file contains 17 adjustable parameters, allowing users to quickly modify neutron wavelength, monochromator crystal type and focusing mode, collimator divergence, detector pixel size and position, among others. These settings facilitate the evaluation of how different configurations affect instrument performance (e.g., flux and resolution).
    
    \item \textbf{Integration with External Simulation Tools} \\
    Since the simulation starts at the entrance of the neutron guide, it is well-suited for integration with neutron source data obtained from MCNP simulations. This allows for realistic modeling of the neutron field and source distribution.
    
    \item \textbf{Potential for GUI Integration} \\
    The system can be extended with a graphical user interface (GUI), for example using Python, to help users select configurations and automatically generate input files. This enhancement would reduce the learning curve and accelerate parameter scans.
    
    \item \textbf{Expandable Simulation Scope} \\
    The input framework can be further extended by incorporating additional McStas components such as bent guides, secondary monochromators, and analyzer crystals, enabling advanced instrument simulation and education-oriented demonstration tools.
\end{enumerate}

\section{Conclusion}

This study presents a modular and customizable McStas-based simulation framework tailored for the preliminary design and performance analysis of thermal neutron diffractometers. The simulation includes major optical components such as collimators, a configurable focusing monochromator, a sample model, and a 2D position-sensitive detector. Key design variables—such as divergence, mosaic spread, crystal orientation, and focusing mode—are fully parameterized.

Simulation results under various geometrical and optical conditions demonstrate that the tool accurately reproduces expected trends in neutron flux and resolution, aligning well with theoretical predictions from analytical models. This confirms the framework's suitability for early-stage instrument development and performance benchmarking.

The architecture is highly flexible: users can simulate a wide range of scenarios by adjusting input parameters without modifying the component logic. It can also be extended for more advanced use cases, including integration with MCNP-derived source terms or a GUI interface for rapid prototyping and instructional use.

To encourage broader adoption and further development,the McStas simulation input files utilized in this study have been compiled and presented in  \hyperref[appendix:B]{\textbf{Appendix B}} for reference and potential adaptation.

In addition, a graphical interface for analytical resolution calculations is currently in its final development phase and will soon be released for non-commercial use in academic and research institutions.

\section*{Declaration of generative AI and AI-assisted technologies in the writing process}

During the preparation of this work, the authors used ChatGPT (OpenAI) to improve the clarity and readability of the English language. After using this tool, the authors carefully reviewed and edited the content to ensure accuracy and take full responsibility for the final version of the manuscript.

\appendix
\clearpage
\section*{Appendix A: McStas Input Parameter Table} 
\label{appendix:A}
\begin{table}[htbp]
\centering
\caption{Adjustable Input Parameters in McStas}
\begin{tabular}{|l|p{9cm}|l|}
\hline
\textbf{Parameter Name} & \textbf{Description} & \textbf{Unit} \\
\hline
I0 & Neutron source intensity & s$^{-1}$ \\
Lambda & Desired neutron wavelength & $10^{-10}$ m \\
Beta & Mosaic spread of the monochromator crystal & arcmin \\
Mono\_d & Lattice spacing of the monochromator crystal & $10^{-10}$ m \\
Mono\_reflect & Reflectivity of the monochromator crystal & -- \\
Focus\_H & Horizontal focusing (0: Off, 1: On) & 0 or 1 \\
Focus\_V & Vertical focusing (0: Off, 1: On) & 0 or 1 \\
Co1 & Collimator 1 divergence angle & arcmin \\
Co2 & Collimator 2 divergence angle & arcmin \\
Co3 & Collimator 3 divergence angle & arcmin \\
L\_MS & Distance from monochromator to sample & m \\
L\_MSlit2 & Distance from monochromator to slit & m \\
sample & Type of sample & -- \\
detector\_bins & Number of detector bins & -- \\
detector\_height & Height of the detector & m \\
m1 & Supermirror parameter of neutron guide 1 & -- \\
m2 & Supermirror parameter of neutron guide 2 & -- \\
\hline
\end{tabular}
\label{tab:input_parameters}
\end{table}

\clearpage
\appendix
\section*{Appendix B: McStas Input Code}
\label{appendix:B}
\addcontentsline{toc}{section}{Appendix B: McStas Input Code}

\lstset{
  basicstyle=\ttfamily\tiny,
  numbers=left,
  numberstyle=\tiny,
  stepnumber=1,
  numbersep=5pt,
  frame=single,
  breaklines=true,
  breakatwhitespace=false,
  postbreak=\mbox{\textcolor{red}{$\hookrightarrow$}\space},
  tabsize=2,
  backgroundcolor=\color{gray!5},
  keywordstyle=\color{blue},
  commentstyle=\color{green!50!black},
  stringstyle=\color{orange},
}

\begin{lstlisting}[language=C, caption={ Monochromator Diffractometer McStas Input Code}]
/*******************************************************************************
*         McStas instrument definition URL=http://www.mcstas.org
*
* Instrument: NARI monochromator Diffractometer
*
* %Identification
* Written by: Chen LiFang
* Date: 5 July 2024
* Origin: NARI
* %INSTRUMENT_SITE: NARI
*
* Simple monochromator Diffractometer for powders (D1A) .
*
* %Description
* Monochromator           
* HOPG 002 DM=3.35 AA
* Ge 111 DM=3.26 AA
* Ge 115 DM=1.0925 AA     
* Ge 113 DM=1.712 AA    
*
* %Parameters
* I0:[/s]                    Intensity of neutron source.
* lambda: [Angs]              Target Wavelength at sample stage. 
* Beta: [minutes of arc]      Mosaic spread of Monochromator.
* Mono_d: [Angs]              d-spacing of monochromator.
* Focus_H:                   Focus function in Horizontal direction. 0/1:closed/opened
* Focus_V:                   Focus function in Vertical direction. 0/1:closed/opened
* Co1: [minutes of arc]       Divergence horizontal angle of collimator 1.
* Co2: [minutes of arc]       Divergence horizontal angle of collimator 2.
* Co3: [minutes of arc]       Divergence horizontal angle of collimator 3.
* L_MS:[m]                   The distance between monochromator and sample stage.
* L_MSlit3:[m]                The distance between monochromator and Slit3.
* string sample: [str]        File name for powder sample description.
* detector_bins:              The range of angle of Banana Detector is 157 degrees. 
*                             The coverage angle of a single detector is 157/detector_bins degrees.
* detector_height:[m]         Height of detector.
* m1:                         m-value of guide1.
* m2:                         m-value of guide2.        
*
* %End
*******************************************************************************/


DEFINE INSTRUMENT NARIDiff (I0=4.26e+010,Lambda=1.54,Beta=30,Mono_d=1.0925, Mono_reflect=0.2, Focus_H=1, Focus_V=1,Co1=0,Co2=0,Co3=60,L_MS=2,L_MSlit3=1.95,string sample="Ag.laz",detector_bins=157,detector_height=0.2,m1=3,m2=1)

/* Version of 20240331  By LiFangChen*/

DECLARE

%{
  double Ki, Co1,Co2 ,Co3,Ein, dEin, sina1, A1, A2, RVN, RHN, L0, L_MS, L_MSlit3,LVS, I0, Lambda,Mono_d, Mono_reflect, Q0,Beta,pixel,detector_height,m1,m2;
  double flag_container,flag_sample,flag_env;
  double verbose=1;
  double Focus_H, Focus_V;
  int detector_bins;
  char str1[60]={"banana, theta limits=[-160 -3], bins="};
  char str2[6];
  char sample1[60];
%}

INITIALIZE

/* E=p^2/2m = (hbar^2)*(Ki^2)/2m =2.07212*Ki^2 p=k * hbar = 1.054573 E -34 * k; m=1.67493 e-27 
   1 meV = 1.60218 e-22 */
%{
#include <stdio.h>
#include <string.h>

Ki = 2*PI/Lambda;
Ein=2.07212*Ki*Ki;
dEin=Ein*0.10;
/*Lc=2*PI/Ki;*/
sina1= PI/Mono_d/Ki;
A1=180.0*asin(sina1)/PI;
A2=2.0*A1;
pixel = (160.0-3.0)/detector_bins;

itoa( detector_bins, str2, 10);
strcat(str1, str2);


/*  RVN   formula corresponds to focusing at source  */
/*  RHN   formula corresponds to focusing at virtual source  */
/*  L_MS  monochromator-sample distance  */
/*  LVS   virtual source - monochromator  */

L0=7; /*source to Guide End*/
LVS=0.5;

if (Focus_H == 1)
    {RHN=2.0*(LVS+L0)*L_MS/sina1/(LVS+L0+L_MS);}

if (Focus_V == 1)
    {RVN=2.0*sina1*(LVS+L0)*L_MS/(LVS+L0+L_MS);}


Q0 = 2*PI/Mono_d;

printf("A1 (deg) A1 = %f\n",A1);
printf("A2 (deg) A2 = %f\n",A2);
printf("Ki = %f\n",Ki); 
printf("Ei = %f\n",Ein);
printf("Q0 = %f\n",Q0); 
printf("Lambda = %f\n",Lambda); 
printf("L0 = %f m\n",L0); 
printf("m1 = %f \n",m1);
printf("m2 = %f \n",m2);
printf("Beta = %f\n",Beta); 
printf("Mono_d = %f\n" ,Mono_d); 
printf("Mono_reflect = %f\n",Mono_reflect);
printf("RH = %f\n",RHN); 
printf("RV = %f\n",RVN);
printf("L_MS = %f\n",L_MS);
printf("Co1 = %f\n",Co1);
printf("Co2 = %f\n",Co2);
printf("Co3 = %f\n",Co3);
printf("pixel = %f degrees\n",pixel);
printf("sample = %s\n",sample);
printf("detector = %s\n",str1);
%}

TRACE

COMPONENT src = Arm()  AT (0,0,0) ABSOLUTE

/* souce parameter */
/*Maxwell Boltzman Neutron Source*/
COMPONENT source1 = Source_gen(
                               yheight  = 0.12,
                               xwidth   = 0.06,
                               dist     = 0.02,
                               focus_aw = 1,
                               focus_ah = 1,
                               E0  = 51,
                               dE  = 50,
                               T1=330.43,I1=I0,
                               verbose  = 1) 
AT (0,0,0) RELATIVE src

/*E_Monitor_1*/
COMPONENT E_Monitor_1 = E_monitor(
                          xmin=-0.30,
                          xmax=0.30,
                          ymin=-0.30,
                          ymax=0.30,
                          nE=100,
                          filename="E_Monitor_1.txt",
                          Emin=0,
                          Emax=100)
AT(0,0,0.00011) RELATIVE src

/*COMPONENT PSD_1 = PSD_monitor(
                          xmin = -0.30, 
                          xmax = 0.30,
                          ymin = -0.30, 
                          ymax = 0.30,
                          nx = 120, 
                          ny = 120,
                          filename = "VS_XY.psd")
AT(0,0,0.00012) RELATIVE src*/


/*Slit 1*/
COMPONENT Slit1=Slit(
                          xmin=-0.03,
                          xmax=0.03,
                          ymin=-0.06,
                          ymax=0.06)
AT (0,0,0.0009) RELATIVE src

/*  Super Mirror Guide  */
COMPONENT guide1=Guide(w1 = 0.06, 
                       h1 = 0.12, 
                       w2 = 0.06, 
                       h2 = 0.12, 
                       l = L0,
                       R0=0.99,
                       m=m1)
AT (0,0,0.001) RELATIVE src

/*Slit 2*/
COMPONENT Slit2=Slit(
                          xmin=-0.03,
                          xmax=0.03,
                          ymin=-0.06,
                          ymax=0.06)
AT (0,0,L0+0.001) RELATIVE src


COMPONENT E_Monitor_2 = E_monitor(
                          xmin=-0.30,
                          xmax=0.30,
                          ymin=-0.30,
                          ymax=0.30,
                          nE=100,
                          filename="E_Monitor_2.txt",
                          Emin=0,
                          Emax=100)
AT(0,0,L0+0.0011) RELATIVE src

COMPONENT PSD_2 = PSD_monitor(
                          xmin = -0.30, 
                          xmax = 0.30,
                          ymin = -0.30, 
                          ymax = 0.30,
                          nx = 120, 
                          ny = 120,
                          filename = "VS_XY.psd")
AT(0,0,L0+0.0011) RELATIVE src


/* Collimator No1 */
COMPONENT Collimator_No1= Collimator_linear(
                          xmin= -0.05, 
                          xmax= 0.05, 
                          ymin= -0.08,
                          ymax= 0.08, 
                          length= 0.15, 
                          divergence= Co1,
                          transmission=0.7)
AT(0,0,L0+0.05) RELATIVE src 

/*  Q = 2Pi/d = 2Pi/PG(002) = 2Pi/3.355 = 1.872783  */

/*Monochromator*/
COMPONENT Mono_Cradle = Arm()
AT (0, 0, L0+LVS) RELATIVE src ROTATED (0, A1, 0) RELATIVE src

COMPONENT Ge115_mono = Monochromator_curved(
                          zwidth=0.02,
                          yheight=0.03,
                          gap=0.0005,
                          NH=5,
                          NV=5,
                          mosaich=Beta,
                          mosaicv=Beta,
                          r0=Mono_reflect,
                          Q=Q0,
                          RV=RVN,
                          RH=RHN)
AT (0,0,0.00001) RELATIVE Mono_Cradle 

/* Monitor at beamstop */
/*COMPONENT Beamstop1 =  Beamstop(xmin=-0.050, 
                                  xmax=0.05, 
                                  ymin=-0.1, 
                                  ymax=0.1)
AT(0,0,LVS+0.5) RELATIVE src

COMPONENT PSD_Beamstop1 = PSD_monitor(
                          xmin = -0.20, 
                          xmax = 0.20,
                          ymin = -0.30, 
                          ymax = 0.30,
                          nx = 120, 
                          ny = 120,
                          filename = "Beamstop_XY_1.psd")
AT(0,0,LVS+0.51) RELATIVE src*/


COMPONENT Mono_Out = Arm()      
AT (0,0,0.00002) RELATIVE Mono_Cradle ROTATED (0, A2, 0) RELATIVE src

COMPONENT guide2=Guide(w1 = 0.07, 
                       h1 = 0.14, 
                       w2 = 0.02, 
                       h2 = 0.03, 
                       l = L_MSlit3-0.15,
                       R0=0.99,
                       m=m2)
AT (0,0,0.15) RELATIVE Mono_Out

/* Collimator No2 */
COMPONENT Collimator_No2= Collimator_linear(
                          xmin= -0.035, 
                          xmax= 0.035, 
                          ymin= -0.045,
                          ymax= 0.045, 
                          length= 0.15, 
                          divergence= Co2,
                          transmission=0.7)
AT(0,0,0.5) RELATIVE Mono_Out

/*Slit 3*/
COMPONENT Slit3=Slit(
                          xmin=-0.01,
                          xmax=0.01,
                          ymin=-0.015,
                          ymax=0.015)
AT (0,0,L_MSlit3) RELATIVE Mono_Out

COMPONENT E_Monitor_3 = E_monitor(
                          xmin=-0.30,
                          xmax=0.30,
                          ymin=-0.30,
                          ymax=0.30,
                          nE=100,
                          filename="E_Monitor_sample_3.txt",
                          Emin=0,
                          Emax=100)
AT(0,0,L_MS-0.002) RELATIVE  Mono_Out

COMPONENT PSD_Sample = PSD_monitor(
                          xmin = -0.12, 
                          xmax = 0.12,
                          ymin = -0.12, 
                          ymax = 0.12,
                          nx = 120, 
                          ny = 120,
                          filename = "Sample_XY.psd")
AT(0,0,L_MS-0.001) RELATIVE Mono_Out


/* sample position ********************************************************** */


SPLIT COMPONENT SamplePos=Arm()
AT (0, 0, L_MS) RELATIVE Mono_Out
EXTEND %{
  flag_container=flag_sample=flag_env=0;
%}

/*COMPONENT Environment_in=PowderN(
  radius = 0.05, yheight = 0.1, thickness=0.002,
  reflections="Al.laz", concentric=1, d_phi=RAD2DEG*atan2(0.5,0.5),
  p_transmit=0.95, p_inc=0, barns=1)
  AT (0, 0, 0) RELATIVE SamplePos
EXTEND %{
  flag_env=SCATTERED;
%}

COMPONENT Container_in=PowderN(radius=0.008/2+1e-4, thickness=1e-4, yheight=0.05,
                       reflections= "V.laz", concentric=1, d_phi=RAD2DEG*atan2(0.5,0.5) ,
                       p_transmit=0.93, p_inc=0.05)
AT (0, 0, 0) RELATIVE SamplePos
EXTEND %{
  flag_container=SCATTERED;
%}  */

COMPONENT Sample=PowderN(reflections = sample,
                         radius = 0.008/2, yheight = 0.03,
                         d_phi=RAD2DEG*atan2(0.5,0.5), p_transmit=0.08, p_inc=0.05)
AT (0, 0, 0) RELATIVE SamplePos
EXTEND %{
  if (SCATTERED)
    flag_sample=SCATTERED;
%}


/*COMPONENT Container_out=COPY(Container_in)(concentric=0)
  AT (0, 0, 0) RELATIVE SamplePos
EXTEND %{
  if (SCATTERED) flag_container=1;
%}

COMPONENT Environment_out=COPY(Environment_in)(concentric=0)
  AT (0, 0, 0) RELATIVE SamplePos
EXTEND %{
  if (SCATTERED) flag_env=1;
%}  */


/* sample position (end) **************************************************** */

COMPONENT Collimator_No3=Collimator_radial(xwidth=0, yheight=.25, length=.2,
                      divergence=Co3,transmission=1, nchan=32,
                      theta_min=-160, theta_max=0, radius=0.75)
AT (0, 0, 0) RELATIVE SamplePos
EXTEND %{
  if (!flag_sample && !flag_container) ABSORB;
%}

/* perfect detector: 1D(theta) */
COMPONENT BananaTheta = Monitor_nD(
    options = str1,
    xwidth = 1.0*2, yheight = detector_height)
  AT (0, 0, 0) RELATIVE SamplePos

/* perfect detector: 2D(theta,y) to see diffraction rings */
COMPONENT BananaPSD = Monitor_nD(
    options = "banana, theta limits=[-160 -5] bins=380, y bins=25",
    xwidth = 1.0*2*1.005, yheight = 1.0)
  AT (0, 0, 0) RELATIVE SamplePos

/*COMPONENT Resolution_monitor = Res_monitor(filename="Output.res",
                       res_sample_comp=RSample,
                       options = "banana, theta limits=[-160 -5] ", 
                       xwidth = 1.0*2*1.001, yheight = detector_height)
AT (0, 0, 0) RELATIVE SamplePos*/

/*COMPONENT Resolution_monitor = Res_monitor(filename="Output.res", 
                                           res_sample_comp=Res_sample, 
                                           xmin=-0.1, xmax=0.1, ymin=-0.1, ymax=0.1)
AT (0, 0, 1.5) RELATIVE SamplePos ROTATED (0, 20, 0) RELATIVE SamplePos*/

COMPONENT Beamstop2 =  Beamstop(xmin=-0.05, 
                                  xmax=0.05, 
                                  ymin=-0.1, 
                                  ymax=0.1)
AT(0,0,1.5) RELATIVE SamplePos

COMPONENT PSD_Beamstop2 = PSD_monitor(
                          xmin = -0.20, 
                          xmax = 0.20,
                          ymin = -0.30, 
                          ymax = 0.30,
                          nx = 120, 
                          ny = 120,
                          filename = "Beamstop_XY_2.psd")
AT(0,0,1.51) RELATIVE SamplePos 


END




\end{lstlisting}

\end{document}